\documentclass[a4paper,11pt]{article}
\usepackage{aprox2009en}
\usepackage[top=3cm,left=3cm,right=2cm,bottom=2cm]{geometry}
\usepackage{authblk}
\usepackage{hyperref}
\usepackage[latin1]{inputenc}
\usepackage[english,brazil]{babel}
\usepackage[T1]{fontenc} 

\title{Facility Leasing with Penalties}

\author[1]{Murilo S. de Lima\thanks{\url{mslima@ime.usp.br} Supported by FAPESP PhD Scholarship Process 2014/18781-1, and CNPq PhD Scholarship Process~142161/2014-4.}}
\author[2]{Mário C. San Felice\thanks{\url{felice@ic.unicamp.br} Supported by CAPES PNPD scholarship 1522390.}}
\author[1]{Orlando Lee\thanks{\url{lee@ic.unicamp.br} Partially supported by Bolsa de Produtividade do CNPq Process 311373/2015-1, and Edital Universal CNPq Process 477692/2012-5.}}
\affil[1]{Institute of Computing, University of Campinas (Unicamp)}
\affil[2]{Institute of Mathematics and Statistics, University of São Paulo (USP)}

\newcommand{\reaisnneg}{\mathbb{R}_+}
\newcommand{\naturais}{\mathbb{N}}

\newcommand{\inteirosnneg}{\mathbb{Z}_+}
\newcommand{\recebe}{\leftarrow}

\newcommand{\nil}{\ensuremath{\mathsf{null}}}

         {\begin{taggednumberedtheorem}{Hypothesis}{#1}%
         }%
         {\end{taggednumberedtheorem}%
         }

\begin{document}

\selectlanguage{english}

\maketitle

\begin{abstract}
In this paper we study the facility leasing problem with penalties. We present a
primal-dual algorithm which is a 3-approximation, based on the algorithm by Nagarajan and
Williamson for the facility leasing problem \cite{nagarajan13leasing} and on the algorithm
by Charikar~{\em et~al.} for the facility location problem with penalties
\cite{charikar01outliers}. 
\end{abstract}

\section{Introduction}

In the facility location problem, one is given a set $F$ of facilities, an opening cost
for each facility, a set $D$ of clients and a metric distance function $d$ between
facilities and clients. The objective is to choose a subset of the facilities to open and
an assignment between clients and facilities, so to minimize the cost of opening the
facilities plus the sum of the distances between each client and the corresponding
assigned facility. This is an NP-hard problem, and it does not have a polynomial-time
algorithm with approximation factor smaller than $1.463$ unless
P~=~NP~\cite{sviridenko02facility}. Currently the best approximation factor is $1.488$,
due to an algorithm by Li~\cite{li13facility}.

In the facility location problem with penalties, we may not assign a client $j$ to a
facility if we choose to pay a penalty $\pi_j$. I.e., we must select a subset of the
facilities to open, and a subset of the clients we assign to open facilities; we are going
to pay the penalties for the remaining of the clients. The cost of a solution is,
therefore, the cost of opening the selected facilities, plus the distance between the
client and its corresponding facility for each assigned client, plus the penalty cost for
each unassigned client. Clearly the facility location problem reduces to this problem if
we set $\pi_j = \infty$ for every client $j$. Currently it is known a
$1.5148$-approximation algorithm for this problem \cite{li15penalties}; however, there is
a simpler 3-approximation algorithm by Charikar~{\em et~al.} \cite{charikar01outliers}.
Also, if the penalties obey a submodular function, then there is a 2-approximation
algorithm~\cite{li15penalties}.

In the facility leasing problem, client requests are distributed along the time, and
instead of opening facilities permanently, we may lease each facility for one of $K$
different durations $\delta_1, \ldots, \delta_K$. The cost for leasing a facility for
$\delta_k$ units of time depends on the facility position, as in the traditional facility
location problem, but also on the leasing type $k$. Additionally, it is reasonable to
suppose that the leasing costs
respect an economics of scale: the leasing cost per unit of time decreases with the
leasing duration, for a fixed facility location. A facility lease may begin at any
moment in the time. Then, we wish to select a set of facility leases that cover the client
requests and minimizes the leasing costs plus the distance between each client and the
facility lease that serves each of its requests. This problem was proposed by Nagarajan
and Williamson, who presented a simple 3-approximation primal-dual algorithm
\cite{nagarajan13leasing}.

In this paper, we study the combination of the previous two problems, which we call the
{\bf facility leasing problem with penalties} ({\sc PFLe}). In this problem, facilities
are leased instead of permanently opened, as in the facility leasing problem, and some
clients may be left unassigned by paying for the penalty cost. We obtain a 3-approximation
algorithm by combining the algorithm by Nagarajan and Williamson for the facility leasing
problem \cite{nagarajan13leasing} and the algorithm by Charikar~{\em et~al.} for the
facility location problem with penalties \cite{charikar01outliers}.

The leveraging scheme by Li~{\em et~al.} \cite{li15penalties} implies that, for any
covering problem with an $\alpha$-approximation algorithm, there is a
$(1-e^{-1/\alpha})^{-1}$-approximation algorithm for the corresponding covering problem
with submodular penalties. Combining this with the algorithm by Nagarajan and Williamson
($\alpha = 3$), one may obtain a $3.5277$-approximation algorithm for the facility leasing
problem with submodular penalties. Note that our algorithm obtains a better approximation
ratio for the linear case.

\section{Notation and Problem Definition}

Let $[K] := \{1, \ldots, K\}$ be the set of lease types.
We denote a facility lease by a triple $f = (p_f, k_f, t_f)$, where $p_f \in V$ is the
point where $f$ is located, $k_f \in [K]$ is the leasing type for $f$, and $t_f \in
\inteirosnneg$ is the instant of time in which the lease for $f$ begins.
We write $\Fcal := F \times [K] \times \inteirosnneg$ so to simplify our notation.

Similarly, we denote a client by a triple $j = (p_j, \pi_j, t_j)$, where $p_j \in V$ is the
point where $j$ is located, $\pi_j \in \reaisnneg$ is the penalty for not assigning a
facility lease to $j$, and $t_j$ is the instant in which $j$ arrives.

In order to simplify our notation, we write~$\delta_f$ instead of $\delta(k_f)$, and
$\gamma_f$ instead of $\gamma(p_f, k_f)$, for a facility lease $f = (p_f, k_f, t_f) \in
\Fcal$.
Also, for $f = (p_f, k_f, t_f) \in \Fcal$ and $j = (p_j, \pi_j, t_j) \in V \times
\reaisnneg \times \inteirosnneg$, we define the distance between $j$ and $f$ to be
\begin{displaymath}
 d(j, f) := \left\{ \begin{array}{ll}
  d(p_j, p_f) & \mbox{if } t_j \in [t_f, t_f + \delta_f), \\
  \infty & \mbox{otherwise.} 
 \end{array}\right.
\end{displaymath}
I.e., the distance between client $j$ and facility lease $f$ is infinity if the facility
lease does not cover~$t_j$.

\begin{problema}{{\sc PFLe}$(V, d, F, K, \gamma, \delta, D)$}
The input consists of a set of points $V$, a distance function $d : V \times V
\mapsto \reaisnneg$ between the points of~$V$ satisfying symmetry and triangle
inequality, a set $F \subseteq V$ of potential facilities, an
integer $K > 0$ that represents the number of lease types, a cost $\gamma(p, k) \in
\reaisnneg$ for leasing facility $p \in F$ with leasing type $k \in [K]$, a function
$\delta : [K] \mapsto \naturais$ that maps each lease type to a length in days, and a
set $D \subseteq V \times \reaisnneg \times \inteirosnneg$ of clients in the form $j =
(p_j, \pi_j, t_j)$. The goal is to find a set $X \subseteq \Fcal := F \times [K] \times
\inteirosnneg$ of facility leases in the form $f = (p_f, k_f, t_f)$, and a function $a : D
\mapsto X \cup \{\nil\}$ that maps each client $j$ to an active facility leasing $f \in X$
such that $t_j \in [t_f, t_f + \delta_f)$ or to $\nil$, so to minimize
 \begin{displaymath}
  \sum_{f \in X} \gamma_f + \sum_{j \in D : a(j) \neq \nil} d(j, a(j)) + \sum_{j \in D :
  a(j) = \nil} \pi_j.
 \end{displaymath}
\end{problema}

\section{Primal-Dual Formulation}

\begin{itemize}
 \item Primal:
 \begin{linearprogram}
 \mbox{minimize}
   &\multicolumn{4}{l}{\sum_{f \in \Fcal} \gamma_f \cdot y_f
    + \sum_{j \in D} \sum_{f \in \Fcal} d(j, f) \cdot x_{jf}
    + \sum_{j \in D} \pi_j \cdot z_j} \\
 \mbox{subject to}
   &x_{jf} & \leq & y_f &\forall f \in \Fcal, j \in D \\
   &\sum_{\substack{f \in \Fcal \\ t_j \in [t_f, t_f 
    + \delta_f)}} x_{fj} + z_j & \geq & 1 &\forall j \in D \\
   & x_{fj}, y_f, z_j & \in & \{0, 1\} & \forall f \in \Fcal, j \in D
 \end{linearprogram}
  (Variable $y_f$ indicates whether facility $f$ was leased, variable $x_{jf}$
  indicates whether client~$j$ was served by facility lease $f$, and variable $z_j$
  indicates whether the algorithm decided to pay the penalty associated with not serving
  $j$ with a facility lease.)
 \item Dual relaxation:
 \begin{linearprogram}
 \mbox{maximize}
   & \multicolumn{4}{l}{\sum_{j \in D} \alpha_j} \\[1.5ex]
 \mbox{subject to}
   & \sum_{j \in D}
    (\alpha_j - d(j, f))_+ & \leq & \gamma_f & \forall f \in \Fcal \\
   & \alpha_j & \leq & \pi_j & \forall j \in D \\
   & \alpha_j & \geq & 0 & \forall j \in D
 \end{linearprogram}
  (Economical interpretation: each client $j$ is willing to pay $\alpha_j$ to connect
  itself to some facility lease. Part of this value covers the distance to the facility;
  the other part is a contribution to pay for leasing the facility. However, the client
  is not willing to pay more than its penalty.)
\end{itemize}

\section{Algorithm}

Our algorithm is based on the algorithm by Nagarajan and Williamson for the facility
leasing problem \cite{nagarajan13leasing}, and on the algorithm by Charikar~{\em et~al.}
for the facility location problem with penalties \cite{charikar01outliers}.
We say that a client $j$ {\bf reaches} a facility lease $f$ if $\alpha_j \geq d(j, f)$.

\begin{pseudocode}{\sc Primal-DualPFLe}$(V, d, F, K, \gamma, \delta, D)$

01\x set $\alpha_j \recebe 0$ for every $j \in D$

02\x $X \recebe \emptyset$, $S \recebe D$

03\x while $S \neq \emptyset$ do

04\xx increase $\alpha_j$ uniformly for every $j \in S$ until

05\xxx (a) $\alpha_j = d(j, f)$ for some $j \in S$ and $f \in X$

06\xx or

07\xxx (b) $\gamma_f = \sum_{j \in D} (\alpha_j - d(j, f))_+$ for some $f \in \Fcal
\setminus X$

08\xx or

09\xxx (c) $\alpha_j = \pi_j$ for some $j \in S$

10\xx $X \recebe X \cup \{f \in \Fcal \setminus X : f \mbox{ satisfies (b)}\}$

11\xx $S \recebe S \setminus \{j \in S : \alpha_j \geq \pi_j \mbox{ or $j$ reaches some }
f \in X\}$

12\x build the graph $G_X$ with

13\xx $V[G_X] \recebe X$

14\xx $E[G_X] \recebe \{ (f, f') : \exists j \in D : \mbox{$j$ reaches both $f$ and
$f'$}\}$

15\x build a maximal independent set $X'$ in $G_X$ greedily in decreasing order of
$\delta$

16\x $\hat{X} \recebe \{(p_f, k_f, t_f - \delta_k), f, (p_f, k_f, t_f + \delta_k):
f \in X'\}$

\pagebreak

17\x for every $j \in D$ do

18\xx if $j$ reaches some $f \in X$ then

19\xxx $a(j) \recebe \arg \min_{f' \in \hat{X}} \{d(j, f')\}$

20\xx else

21\xxx $a(j) \recebe \nil$

22\x return $(\hat{X}, a)$

\end{pseudocode}

The algorithm maintains a dual variable $\alpha_j$ for each client $j \in D$, a set $X$ of
temporarily leased facilities, and
a set $S$ of the clients whose dual variable still is being increased, which is
initialized with the whole set of clients $D$. The increasing pauses when either: (a) a
client reaches an already temporarily leased facility, (b) the sum of the contributions
towards a
facility lease pays for its cost or (c) the dual variable reaches the penalty cost for
some client. We then add to $X$ the facilities that reach condition (b).
Also, we remove from $S$ the clients that reach some temporarily leased facility or whose
dual variable pays for the penalty cost, and then proceed the
increasing of the remaining dual variables until $S$ becomes empty.

After that initial phase, we build an interference graph $G_X$ between the facility
leases in $X$. Graph $G_X$ has vertex set $X$ and has an edge between facilities $f$ and
$f'$ if there is some client that reaches both $f$ and $f'$. Then, we order set $X$ by
decreasing order of lease duration and build a maximal independent set $X'$ in a greedily
manner; i.e., we visit set $X$ in that order and add a facility $f$ to $X'$ if there is
no other facility lease $f' \in X'$ reached by some client that reaches~$f$. Thus $X'$
satisfies the following properties:
\begin{enumerate}
 \item Every client reaches at most one facility lease in $X'$;
 \item If facility leases $f$ and $f'$ in $X$ are reached by the same client $j$, and if
 $f' \in X'$, then $\delta_f \leq \delta_{f'}$.
\end{enumerate}

However, note that there may be some client $j$ that reaches some $f$ in $X$ but is not
covered by any facility lease in $X'$. But then remember that some $f' \in X'$ shares a
reaching client $j'$ with~$f$, thus $\delta_f \leq \delta_{f'}$ and the intervals covered
by facility leases $f$ and $f'$ overlap. Then, since we buy $\hat{X}$, which has
three copies of $f'$, beginning at instants $t_{f'} - \delta_{f'}$, $t_{f'}$ and $t_{f'} +
\delta_{f'}$, we have that the interval formed by those three facilities, which is
$[t_{f'} - \delta_{f'}, t_{f'} + 2\delta_{f'})$, is a superset of interval $[t_f, t_f +
\delta_f)$, and therefore one of them covers $t_j$.

Finally, if some client $j$ does not reach any facility lease in $X$, then its dual
variable pays for its penalty and we set $a(j)$ to $\nil$.

Also, note that, although the number of potential facility leases is infinite, the
algorithm may be implemented in finite time, which is also polynomial in the input size:
it is enough to consider, for every facility point, a lease beginning at each instant in
which we have a client request.

\section{Analysis}

In this section we analyze the approximation factor of algorithm {\sc Primal-DualPFLe}.

First note that, since the conditions (a), (b) and (c) correspond to constraints of the
relaxation of the dual program, we have that $\alpha$ is a feasible dual solution.
Therefore, by weak duality, we have that
\begin{displaymath}
 \sum_{j \in D} \alpha_j \leq \opt(V, d, F, K, \gamma, \delta, D).
\end{displaymath}

We will show, then, that the cost of the primal solution $(\hat{X}, a)$ returned by the
algorithm is at most 3 times the cost of the dual solution, and thus our algorithm is
a 3-approximation to problem {\sc PFLe}.

\pagebreak

For every client $j \in D$, we define numbers $\alpha^C_j$, $\alpha^F_j$, and $\alpha^P_j$
in the following manner:
\begin{enumerate}
 \item If $j$ reaches some $f \in X'$, then let
 \begin{displaymath}
  \alpha^C_j := d(j, f), \quad \alpha^F_j := \alpha_j - d(f, j), \quad \alpha^P_j := 0 ;
 \end{displaymath}
 \item If $j$ does not reach any facility lease in $X'$ but reaches some $f \in X$, then
 we let
 \begin{displaymath}
  \alpha^C_j := \alpha_j, \quad \alpha^F_j := 0, \quad \alpha^P_j := 0 ;
 \end{displaymath}
 \item Finally, if $j$ does not reach any facility lease in $X$, then we let
 \begin{displaymath}
  \alpha^C_j := 0, \quad \alpha^F_j := 0, \quad \alpha^P_j := \alpha_j.
 \end{displaymath}
\end{enumerate}
Note that, either case, we have that
\begin{displaymath}
 \alpha_j = \alpha^C_j + \alpha^F_j + \alpha^P_j.
\end{displaymath}

Now, first let us bound the facility leasing cost. Note that, by construction, we have
that, for every $f \in X'$, every client that reaches $f$ reaches only $f$ in $X'$. Also,
by case (b) of the algorithm, the leasing cost of $f$ is totally paid by contributions
from clients that reach $f$. Therefore, we have that
\begin{displaymath}
 \sum_{f \in X'} \gamma_f = \sum_{j \in D} \alpha^F_j.
\end{displaymath}
Since $\hat{X}$, which is the set of facility leases actually bought by the algorithm,
consists of three copies of each facility lease in $X'$, we have that
\begin{displaymath}
 \sum_{f \in \hat{X}} \gamma_f \leq 3 \cdot \sum_{j \in D} \alpha^F_j.
\end{displaymath}

Now we bound the penalty cost. We have that a client $j$ has $a(j)$ set to $\nil$ if and
only if it does not reach any facility lease in $X$, and then $\alpha_j = \alpha^P_j$.
Also, due to case (c) of the algorithm, we have that $\alpha_j = \pi_j$. Thus, it is
straightforward to conclude that
\begin{displaymath}
 \sum_{j \in D : a(j) = \nil} \pi_j = \sum_{j \in D} \alpha^P_j.
\end{displaymath}

Finally, we have to bound the client connection cost. Let $D_{X'}$ be the set of clients
that reach some facility in $X'$. Note that those clients are connected to the closest
facility lease in $\hat{X}$. Since every such client $j$ reaches some $f \in X'$, we have
that
\begin{displaymath}
 d(j, a(j)) \leq d(j, f) = \alpha^C_j.
\end{displaymath}
Now let $j$ be some client that reaches some $f \in X$ but does not reach any facility
lease in $X'$. There must be some $f' \in X'$ and some $j'$ that reaches both $f$ and $f'$,
by construction of $X'$. But then we must have that $\alpha_j \geq \alpha_{j'}$, since when
$\alpha_{j'}$ stopped increasing it reached both $f$ and~$f'$, and $\alpha_j$ reached $f$
when it stopped increasing. Then, since $j'$ reaches both $f$ and $f'$, we have that
\begin{displaymath}
 \alpha_{j'} \geq d(j', f) \quad \mbox{and} \quad \alpha_{j'} \geq d(j', f').
\end{displaymath}
Since one of the three copies of $f'$ in $\hat{X}$ must cover $t_j$, by triangle inequality,
we have that
\begin{displaymath}
 d(j, a(j)) \leq d(j, f') \leq d(j, f) + d(j', f) + d(j', f') \leq \alpha_j + \alpha_{j'} +
\alpha_{j'} \leq 3 \cdot \alpha_j = 3 \cdot \alpha^C_j.
\end{displaymath}

Summing up the previous inequalities, we have that
\begin{displaymath}
 \sum_{f \in \hat{X}} \gamma_f + \sum_{j \in D : a(j) \neq \nil} d(j, a(j)) + \sum_{j \in
 D : a(j) = \nil} \pi_j \leq 3 \cdot \sum_{j \in D} \alpha_j \leq 3 \cdot \opt(V, d, F, K,
 \gamma, \delta, D),
\end{displaymath}
and we conclude the following theorem.

\begin{teorema}{}
 Algorithm {\sc Primal-DualPFLe} is a 3-approximation.
\end{teorema}

\bibliographystyle{alpha}
\bibliography{pfle}

\end{document}